\documentclass{PoS}

\newcommand{\bea}{\begin{eqnarray}}
\newcommand{\eea}{\end{eqnarray}}
\newcommand{\be}{\begin{equation}}
\newcommand{\ee}{\end{equation}}
\newcommand{\lamqcd}{\Lambda_\mathrm{QCD}}
\newcommand{\cO}{{\cal O}}

\newcommand{\mh}{{\mu_b}}
\newcommand{\mhc}{{\mu_\mathrm{hc}}}
\newcommand{\ti}{{T^{\rm I}_i}}
\newcommand{\tii}{{T^{\rm II}_i}}

\newcommand{\TII}{{T^{\rm II}}}
\newcommand{\HII}{{H^{\rm II}}}
\newcommand{\sceti}{{SCET$_{\mathrm{I}}$}}
\newcommand{\scetii}{{SCET$_{\mathrm{II}}$}}
\newcommand{\xib}{{\bar \xi}}
\newcommand{\chib}{{\bar \chi}}
\newcommand{\npslash}{{n \!\!\!\! /}_+ }
\newcommand{\nmslash}{{n \!\!\!\! /}_- }
\newcommand{\nph}{{\frac{\npslash}{2}}}
\newcommand{\nmh}{{\frac{\nmslash}{2}}}

\newcommand{\Dpc}{\not\!\! D_{\perp c}}
\newcommand{\OI}{{O^{\mathrm{I}}}}
\newcommand{\OII}{{O^{\mathrm{II}}}}
\newcommand{\hf}{{\frac{1}{2}}}

\title{Hard spectator-scattering in $B\to\pi \pi$ decays
at NNLO\footnote{PITHA 05/18}}

\ShortTitle{Hard spectator-scattering in $B\to\pi \pi$ 
decays at NNLO}

\author{Martin Beneke\\
        Institut f\"ur Theoretische Physik E, RWTH Aachen\\
        D - 52056 Aachen, Germany\\
        E-mail: \email{mbeneke@physik.rwth-aachen.de}}

\author{\speaker{Sebastian J{\"a}ger}\\
        Institut f\"ur Theoretische Physik E, RWTH Aachen\\
        D - 52056 Aachen, Germany\\
        E-mail: \email{sjaeger@physik.rwth-aachen.de}}

\abstract{We compute the 1-loop (NNLO) corrections to hard 
  spectator-scattering in tree-dominated hadronic $B$ decays. Depending 
  on the values of hadronic input parameters the corrections 
  are shown to have a significant impact on the $B\to \pi\pi$ 
  branching fractions.}

\FullConference{
  International Europhysics Conference on High Energy Physics\\
  July 21st - 27th 2005\\
  Lisboa, Portugal}

\begin{document}

\section{Introduction}
\noindent 
Hadronic $B$-decays into two light mesons are a rich source of
information about the flavour structure of the Standard Model (CKM matrix)
and of its possible extensions. On the theoretical side the task of 
relating the fundamental parameters to the large number of
branching ratios and CP asymmetries on which experimental data is available
is complicated by strong interaction effects.
In recent years, these have become more tractable through the development
of the QCD factorization (QCDF) formalism~\cite{BBNS} and the
soft-collinear effective theory (SCET)~\cite{scetmomspace,scetposspace}.

At leading power in an expansion in $\lamqcd/m_b$ the matrix elements
of operators in the weak effective Hamiltonian obey a factorization
formula~\cite{BBNS}
\bea                           
  \langle M_1 M_2 | O_i | \bar B \rangle
        &=& f_+^{B\to M_1}(0) \int {\rm d}u\, \ti(u)\, \phi_{M_2}(u)
               + \int {\rm d} \omega\, {\rm d}u\, {\rm d} v\,
                \tii(\omega, v, u) \phi_{M_1}(v) \phi_{M_2}(u) .
\label{eq:factform}
\eea
The hard-scattering kernels $\ti,\tii$ are perturbatively calculable
in the strong coupling, while the form factors $f_+$ and 
light-cone distribution amplitudes (LCDAs) $\phi$ encapsulate 
nonperturbative
properties of the initial and final state particles. Both kernels
are currently known to $\cO(\alpha_s)$.

Within the set of hadronic final states, the $\pi\pi$ system is
special. Isospin symmetry allows to extract
the strong amplitudes if the CKM angle $\gamma$ is known. 
Consequently it can be considered a test case for a theoretical 
description of QCD dynamics
such as QCDF. Indeed, the naive factorization result predicts a
$1/N_c^2 \sim 0.1$ colour suppression of $BR(\bar B_d \to \pi^0 \pi^0)$,
stronger than what is observed in experiment. 
At $\cO(\alpha_s)$ the kernel $\ti$ for the colour-suppressed tree 
amplitude $\alpha_2(\pi\pi)$ receives corrections proportional to
the large Wilson coefficient $C_1$ that nearly cancel the LO
contribution, increasing its sensitivity to the spectator-scattering 
term $\tii$. This motivates a computation of
the 1-loop current-current contributions (NNLO,
$\cO(\alpha_s^2)$) to $\tii$ for the
tree operators $Q_1$ and $Q_2$, on which we report below. 
The 1-loop correction also introduces a new source of 
strong phases (from spectator-scattering) not present in 
the NLO (tree) contribution to  $\tii$. 

\section{Matching calculation}

\noindent
Within the SCET framework, the spectator-scattering kernel $\tii$ arises
through two consecutive matching steps (e.g., third paper 
of~\cite{scetmomspace}). The \sceti\ operators relevant to this
calculation are 
\bea
  \OI(t) &=& \chib (t n_-) \nmh (1 - \gamma_5) \chi(0) \;
        \big[ \tilde C_A \xib(0) \npslash (1 - \gamma_5) h_v(0) 
\nonumber \\*
        && \qquad \qquad \qquad \qquad \qquad
                - \frac{1}{m_b} \int {\rm d}s\, \tilde C_B(s)
                \xib(0) \npslash i \!\Dpc (s n_+)
                (1 + \gamma_5) h_v(0) \big] ,
\nonumber \\
  \OII(t,s) &=& \chib (t n_-) \nmh (1 - \gamma_5) \chi(0) \;
        \frac{1}{m_b} \xib(0) \nph i \!\Dpc (s n_+)
                (1 + \gamma_5) h_v(0) .
\eea
The operator $\OI$ is associated with the first term on the 
right-hand side of (\ref{eq:factform}). It 
includes the matching coefficients $\tilde C_A$, 
$\tilde C_B$ such that its matrix element
reproduces the full form factor $f_+$. Our new result refers to the 
1-loop matching coefficient of $\OII(t,s)$. The matrix element 
of this operator factorizes into the LCDA $\phi_{M_2}(u)$ 
and the matrix element 
\be
\frac{1}{m_b} \langle M_1| \xib(0) \nph i \!\Dpc (s n_+) 
(1 + \gamma_5) h_v(0) |\bar{B}\rangle.
\ee
Performing a second matching to \scetii, this matrix element
factorizes into $J_\parallel * \phi_{B} * \phi_{M_1}$~\cite{Beneke:2003pa}
such that
\be                     \label{eq:tiifact}
  \TII(\omega,u,v) = \int {\rm d}v' \HII(u,v') J_\parallel(1-v'; \omega, v) .
\ee
The jet kernel $J_\parallel$ has already been computed in~\cite{jet}. 
To obtain the kernel $\HII$, we computed the amplitude for
$b \to q_c A_{\perp c} \bar q_{\bar c} q_{\bar c}$ up to one 
loop, where the $b$ quark
is at rest and the other partons are collinear with
the light mesons, both in QCD and in \sceti. On the \sceti\ side
the contribution of the operator
$\OI$ serves to eliminate the ``factorizable'' QCD diagrams and provides
a further subtraction term proportional to the known 1-loop kernel
$T^{\rm I(1)}$. The renormalized 1-loop matrix element of $\OII$
is given in terms of the Brodsky-Lepage kernel
and the anomalous dimension kernel of the subleading \sceti\ current.
Here care must be taken to ensure the vanishing of evanescent operator
matrix elements. After these subtractions, the remaining difference 
between the QCD and \sceti\ amplitudes is infrared
finite and is absorbed into an $\cO(\alpha_s)$ correction
$H^{\rm II(1)}$. This together with $J_\parallel$ provides a complete
result for spectator-scattering at NNLO.

To incorporate our results into the colour-suppressed tree amplitude 
we write, following the notation of~\cite{BN},
\bea                          
  \alpha_2(M_1 M_2)  &=&
        C_2 + \frac{C_1}{N_c}
        + \frac{C_1}{N_c} \frac{\alpha_s(\mh) C_F}{4\pi} V_2(\pi)
        + \frac{\pi \alpha_s(\mhc) C_F}{N_c}
                \bigg( \frac{C_1}{N_c}  \big[
                        H_2^\mathrm{tw2}(\pi\pi) I_\parallel
                        + H_2^\mathrm{tw3}(\pi\pi) \big]
\nonumber \\
&& \hspace{0cm}
        + \frac{\alpha_s(\mhc)}{4\pi N_c} 
          \frac{9 f_{M_1}\hat f_B(\mhc)}{m_b f_+(0)\lambda_B(\mhc)}
          \big[ C_1 R_1 + C_2 R_2 \big] \bigg).
  \label{eq:alphaformula}
\eea
An analogous equation with $C_1\leftrightarrow C_2$ 
holds for the colour-allowed tree amplitude $\alpha_1$. $I_\parallel$ 
is the jet function correction [second paper of \cite{jet},
Eq.~(96)], while for the new contributions $R_1$ and $R_2$, 
we find, using asymptotic LCDAs ($x_b=m_b^2/\mhc^{\!2}$), 
\bea
        R_1 &=& C_F \bigg(- \hf \ln^2 x_b+ \hf \ln x_b + 
                  \frac{9}{2} - \frac{3 \pi^2}{4}
                  + 2 i \pi \bigg) 
\nonumber\\
&&      + \bigg(\!C_F- \frac{C_A}{2}\bigg)
                        \bigg( \bigg[2 
                + \frac{2 \pi^2}{3}\bigg] \ln x_b 
                -\frac{74}{5} - 2 \pi^2 + \frac{32}{5} \zeta(3)
                - \bigg(1 + \frac{2 \pi^2}{5}\bigg) i \pi \bigg),
\nonumber \\
        R_2 &=& 3 \ln x_b -\frac{163}{20} + \frac{\pi^2}{3} 
                - \frac{14}{5} \zeta(3)
                + \bigg(\! -3 + \frac{2 \pi^2}{15} \bigg) i \pi.
\eea
The finiteness of these expressions proves factorization of
spectator-scattering at $\cO(\alpha_s^2)$.

\section{Phenomenological implications}

\noindent
Numerically, with input parameters defined in \cite{BN}, 
we obtain 
\begin{eqnarray}
\alpha_2(\pi\pi) &=& 0.17 - [0.17+0.08i\,]_{V_2} + 
\left\{\begin{array}{lc}
[0.10\cdot (1.32+0.40 \,i\,)]_{H_2^{\rm tw2}\,I_\parallel+R_{1,2}} + 
[0.06] _{H_2^{\rm tw3}} & \quad \mbox{(default)}
\\
{[0.28\cdot (1.61+0.49 \,i\,)]}_{H_2^{\rm tw2}\,I_\parallel+R_{1,2}} + 
{[0.17]} _{H_2^{\rm tw3}} 
& \quad \mbox{(S4)}
\end{array}
\right.
\nonumber\\[0.2cm]
&=&
\left\{\begin{array}{lc}
0.20 \,(0.18) - 0.04\,(-0.08)\,i & \qquad \mbox{(default)}
\\ 
0.62 \,(0.47) + 0.05\,(-0.08)\,i & \qquad \mbox{(S4)}
\end{array}
\right.
\end{eqnarray}
The various terms and factors correspond to those in 
(\ref{eq:alphaformula}) and we show the numbers for the default 
parameter set and the set S4 that provides a better overall 
description of hadronic two-body modes. The numbers in parentheses in the
second line give the result from \cite{BN} for comparison. 
Depending on parameters the 1-loop correction 
to spectator-scattering can result in a significant enhancement 
of the real part of the amplitude, predominantly from the jet
function correction [cf. second paper of \cite{jet},
Eq.~(132)], and a substantial correction to the imaginary
part (strong phase) from the new hard-matching correction. 
We confront the old NLO and our new (partial) NNLO results 
with the experimental data
on the three branching ratios (Table~\ref{tab:br}), and  
observe that that the agreement is rather good with set S4. 
More details on the numerical analysis and the (substantial) theoretical 
uncertainties can be found in \cite{inprep}. The numbers for the 
branching fractions in the Table should be considered as preliminary, 
since the NNLO correction to spectator-scattering is still missing 
for the penguin amplitudes (see \cite{Li:2005wx} for a related calculation). 
For this reason we do not discuss CP asymmetries, which we expect to 
be affected by the spectator-scattering phase. 

\renewcommand{\arraystretch}{1.4}
\begin{table}
\begin{center}
\renewcommand{\baselinestretch}{2}
\begin{tabular}{|c||cc|cc||c|}\hline
& \multicolumn{2}{c|}{default} & \multicolumn{2}{c||}{S4}
& Expt.~\cite{HFAG} \\[-0.2cm]
 Mode & NLO & NNLO & NLO & NNLO 
& \\
\hline
$\alpha_1(\pi\pi)$ & 
$0.99+0.02i$ & $0.99-0.01i$ & $0.88+0.02i$ & $0.86-0.07i$ & --\\
\hline
$\alpha_2(\pi\pi)$ & 
$0.18-0.08i$ & $0.20-0.04i$ & $0.47-0.08i$ & $0.62+0.05i$& -- \\
\hline
$\mbox{Br}(\bar{B}^0\to \pi^+\pi^-)$ & 
$8.86$ & $8.89$ & $5.17$ &  $5.10$ & $5.0 \pm 0.4$\\
\hline
$\mbox{Br}(\bar{B}^0\to \pi^0\pi^0)$ & 
$0.35$ & $0.37$ & $0.70$ & $1.11$  & $1.5 \pm 0.3$\\
\hline
 $\mbox{Br}(B^-\to \pi^-\pi^0)$& 
$6.03$ & $6.30$ & $5.07$ & $6.10$ & $5.5 \pm 0.6$ \\
\hline
\end{tabular}
\end{center}
\caption{Tree amplitude coefficients $\alpha_1$ and $\alpha_2$, and the 
CP-averaged $\pi\pi$ branching ratios in units of $10^{-6}$ in 
the default and S4 scenarios of \cite{BN}, showing the 
effect of the new NNLO correction. \label{tab:br}}
\end{table}


\begin{thebibliography}{99}
\bibitem{BBNS}
%\cite{Beneke:1999br}
%\bibitem{Beneke:1999br}
  M.~Beneke, G.~Buchalla, M.~Neubert and C.~T.~Sachrajda,
  %``{QCD} factorization for B $\to$ pi pi decays: Strong phases and CP
  %violation in the heavy quark limit,''
  Phys.\ Rev.\ Lett.\  {\bf 83}, 1914 (1999);
%  [hep-ph/9905312];
  %%CITATION = HEP-PH 9905312;%%
%\cite{Beneke:2000ry}
%\bibitem{Beneke:2000ry}
%  M.~Beneke, G.~Buchalla, M.~Neubert and C.~T.~Sachrajda,
  %``QCD factorization for exclusive, non-leptonic B meson decays: General
  %arguments and the case of heavy-light final states,''
  Nucl.\ Phys.\ B {\bf 591}, 313 (2000);
%  [hep-ph/0006124];
  %%CITATION = HEP-PH 0006124;%%
%\cite{Beneke:2001ev}
%\bibitem{Beneke:2001ev}
%  M.~Beneke, G.~Buchalla, M.~Neubert and C.~T.~Sachrajda,
  %``QCD factorization in B $\to$ pi K, pi pi decays and extraction of
  %Wolfenstein parameters,''
  Nucl.\ Phys.\ B {\bf 606}, 245 (2001).
%  [hep-ph/0104110].
  %%CITATION = HEP-PH 0104110;%%
% SCET, momentum space
\bibitem{scetmomspace}
  C.~W.~Bauer, S.~Fleming, D.~Pirjol and I.~W.~Stewart,
  %``An effective field theory for collinear and soft gluons: Heavy to light
  %decays,''
  Phys.\ Rev.\ D {\bf 63}, 114020 (2001);
%  [hep-ph/0011336];
  %%CITATION = HEP-PH 0011336;%%
%\cite{Bauer:2001yt}
%\bibitem{Bauer:2001yt}
  C.~W.~Bauer, D.~Pirjol and I.~W.~Stewart,
  %``Soft-collinear factorization in effective field theory,''
  Phys.\ Rev.\ D {\bf 65}, 054022 (2002);
%  [hep-ph/0109045].
  %%CITATION = HEP-PH 0109045;%%
%\cite{Bauer:2004tj}
%\bibitem{Bauer:2004tj}
  C.~W.~Bauer, D.~Pirjol, I.~Z.~Rothstein and I.~W.~Stewart,
  %``B $\to$ M(1) M(2): Factorization, charming penguins, strong phases, and
  %polarization,''
  Phys.\ Rev.\ D {\bf 70} (2004) 054015.
%  [arXiv:hep-ph/0401188].
  %%CITATION = HEP-PH 0401188;%%

% SCET, position space
\bibitem{scetposspace}
%\cite{Beneke:2002ph}
%\bibitem{Beneke:2002ph}
  M.~Beneke, A.~P.~Chapovsky, M.~Diehl and Th.~Feldmann,
 %``Soft-collinear effective theory and heavy-to-light currents beyond  leading
  %power,''
  Nucl.\ Phys.\ B {\bf 643}, 431 (2002);
%  [hep-ph/0206152];
  %%CITATION = HEP-PH 0206152;%%
%\cite{Beneke:2002ni}
%\bibitem{Beneke:2002ni}
  M.~Beneke and Th.~Feldmann,
  %``Multipole-expanded soft-collinear effective theory with non-abelian gauge
  %symmetry,''
  Phys.\ Lett.\ B {\bf 553}, 267 (2003).
%  [hep-ph/0211358].
  %%CITATION = HEP-PH 0211358;%%
%\cite{Beneke:2003pa}
\bibitem{Beneke:2003pa}
  M.~Beneke and Th.~Feldmann,
  %``Factorization of heavy-to-light form factors in soft-collinear  effective
  %theory,''
  Nucl.\ Phys.\ B {\bf 685} (2004) 249.
  %%CITATION = HEP-PH 0311335;%%

\bibitem{jet}
%\cite{Hill:2004if}
%\bibitem{Hill:2004if}
  R.~J.~Hill, T.~Becher, S.~J.~Lee and M.~Neubert,
  %``Sudakov resummation for subleading SCET currents and heavy-to-light  form
  %factors,''
  JHEP {\bf 0407} (2004) 081;
%  [arXiv:hep-ph/0404217].
  %%CITATION = HEP-PH 0404217;%%
%\cite{Beneke:2005gs}
%\bibitem{BY}
  M.~Beneke and D.~Yang,
  %``Heavy-to-light B meson form factors at large recoil energy: Spectator
  %scattering corrections,''
  to appear in Nucl. Phys. B [hep-ph/0508250];
  %%CITATION = HEP-PH 0508250;%%
%\cite{Kirilin:2005xz}
%\bibitem{Kirilin:2005xz}
  G.~G.~Kirilin,
  %``Loop corrections to the form factors in B $\to$ pi l nu decay,''
  [hep-ph/0508235].
  %%CITATION = HEP-PH 0508235;%%
%\cite{Beneke:2003zv}
%\bibitem{Beneke:2003zv}
\bibitem{BN}
  M.~Beneke and M.~Neubert,
  %``QCD factorization for B $\to$ P P and B $\to$ P V decays,''
  Nucl.\ Phys.\ B {\bf 675} (2003) 333.
%  [hep-ph/0308039].
  %%CITATION = HEP-PH 0308039;%%
\bibitem{HFAG}
  Heavy Flavour Averaging Group, 
  http://www.slac.stanford.edu/xorg/hfag/index.html
\bibitem{inprep}
M.~Beneke and S.~J\"ager, in preparation.
%\cite{Li:2005wx}
\bibitem{Li:2005wx}
  X.~q.~Li and Y.~d.~Yang,
 %``Revisiting B $\to$ pi pi, pi K decays in QCD factorization approach,''
  Phys.\ Rev.\ D {\bf 72} (2005) 074007.
%  [hep-ph/0508079].
  %%CITATION = HEP-PH 0508079;%%

\end{thebibliography}
\end{document}